\documentclass[onecolumn,showpacs,floatfix,nofootinbib,11pt]{revtex4}
\usepackage{amssymb,amsmath}
\usepackage{subfigure}
\usepackage[dvips]{graphics,color}
\usepackage{epsfig}\usepackage{float}

\newcommand{\II}{\mathbb{I}}

\newcommand{\pa}{\partial}

\newcommand{\n}{\nonumber\\}

\newcommand{\LLL}{\mathcal{L}}
\newcommand{\tth}{\Theta}

\newcommand{\bec}{\begin{center}}
\newcommand{\eec}{\end{center}}

\newcommand{\bea}{\begin{array}}
\newcommand{\ear}{\end{array}}

\newcommand{\bfr}{\begin{flushright}}

\newcommand{\efr}{\end{flushright}}
\newcommand{\noi}{\noindent}

\newcommand{\RR}{\mathbb{R}}

\newcommand{\bege}{\begin{equation}}
\newcommand{\enge}{\end{equation}}

\newcommand{\g}{\gamma}

\newcommand{\beq}{\begin{eqnarray}}\newcommand{\benu}{\begin{enumerate}}\newcommand{\enu}{\end{enumerate}}
\newcommand{\eeq}{\end{eqnarray}}

\newcommand{\CC}{\mathbb{C}}

\newcommand{\lam}{\lambda}

\newcommand{\OO}{\mathbb{O}}

\newcommand{\mmu}{{\mathfrak{u}}}

\newcommand{\bx}{\begin{pmatrix}}
\newcommand{\ex}{\end{pmatrix}}
\newcommand{\ka}{\kappa}

\newcommand{\dual}[1]{\overset{{}^{{}^{\boldsymbol{\neg}}}}{\smash[t]{#1}}}
\newcommand{\du}{\dual{\lam}\lam}
\usepackage{amsfonts}

\begin{document}

\title{From Dirac Action to ELKO Action}

\author{J. M. Hoff da Silva}
\email{hoff@ift.unesp.br} \affiliation{Instituto de F\'{\i}sica
Te\'orica, Universidade Estadual Paulista, Rua Pamplona 145,
01405-900 S\~ao Paulo, SP, Brazil}
\author{Rold\~ao da Rocha}
\email{roldao.rocha@ufabc.edu.br} \affiliation{ Centro de
Matem\'atica, Computa\c c\~ao e Cogni\c c\~ao, Universidade
Federal do ABC, 09210-170, Santo Andr\'e, SP, Brazil}


\begin{abstract}
A fundamental action, representing a mass dimension-transmuting
operator between Dirac and ELKO spinor fields, is performed on the
Dirac Lagrangian, in order to lead it into the ELKO Lagrangian.
Such a dynamical transformation can be seen as a natural extension
of the Standard Model  that incorporates dark matter fields. The
action of the  mass dimension-transmuting operator on a Dirac
spinor field, that defines and introduces such a mapping, is shown
to be a composition of the Dirac operator and the non-unitary
transformation that maps Dirac spinor fields into ELKO spinor
fields, defined in \cite{osmano}. This paper gives allowance for
ELKO, as a candidate to describe dark matter, to be incorporated
in the Standard Model. It is intended to present for the first
time, up to our knowledge, the dynamical character of a mapping
between Dirac and ELKO spinor fields, transmuting the mass
dimension of spin one-half fermionic fields from 3/2 to 1 and from
1 to 3/2.

\end{abstract}
\maketitle

\newpage

\section{Introduction}

ELKO spinor fields\footnote{The acronym for \emph{Eigenspinoren
des Ladungskonjugationsoperators} or Dual-helicity eigenspinors of
the charge conjugation operator \cite{allu}.} are unexpected spin
one-half --- presenting mass dimension 1
--- matter fields, which belong to a non-standard Wigner class
\cite{allu,alu2}, and are obtained from a complete set of
dual-helicity eigenspinors of the charge conjugation operator. Due
to the unusual mass dimension, ELKO spinor fields interact in few
possibilities with the Standard Model particles, which instigates
it to be a prime candidate to describe dark matter\footnote{Other
motivations for the ELKO to be a prime candidate to describe dark
matter can be seen in, e.g., \cite{allu,alu2}.}. Indeed,  the new
matter fields --- constructed via ELKO \cite{gau} --- are dark
with respect to the matter and gauge fields of the Standard Model
(SM), interacting only with gravity and the Higgs boson
\cite{allu,alu2}. Moreover, it is essential  to try to incorporate
ELKO spinor fields in some extension of the SM, identifying  new
fields to dark matter and suggesting how  the dark matter sector
Lagrangian density arises from a mass dimension-transmuting
symmetry.

It was realized in \cite{osmano} that there exists a non-unitary
transformation that can map Dirac spinor fields into ELKO spinor
fields, transmuting the mass dimension of spin one-half fermionic
fields from 3/2 to 1 --- and from 1 to 3/2. The mapping was
obtained using algebraic constraints of the Lounesto spinor fields
algebraic classification, based upon the bilinear covariants
\cite{wal,lou1,lou2}, and it was shown that there always exists an
invertible operator $M$ such that $M\psi=\lam$ --- and obviously
$\psi=M^{-1}\lam$ --- where $\psi$ is a Dirac spinor field and
$\lam$ denotes an ELKO. Also $\bar\psi=\lam^\dagger
(M^{-1})^\dagger\g^0$ \cite{osmano}.

Now we additionaly consider the possibility of incorporating the
dynamics of ELKO spinor fields, extending the SM in order to
accomplish the dynamical, as well the not less fundamental,
algebraic, topological and geometric properties, associated with
ELKO. We also emphasize that all the formalism in this paper is
exhibited from a classical field theoretical point of view.

This paper is organized as follows: in Section II, ELKO spinor
fields are briefly recalled, and in Section III the operator which
maps Dirac spinors fields in ELKO, generically called $M$ here, is
recalled. For the sake of completeness, there is also an Appendix
explaining further topics on the construction of such an operator.
In Section IV the mass dimension-transmuting operator $\Theta$ ---
related to the mapping of the Dirac Lagrangian into the ELKO
Lagrangian --- is introduced, where its action on spinor fields is
shown to be the Dirac operator composed with the operator that
leads Dirac spinor fields to ELKO. The ELKO Lagrangian is then
obtained when a particular class of the operator $\tth$ is
considered, apart from a surface term that can be dismissed when
suitable boundary conditions are chosen. In the final Section we
conclude summarizing our results, remarking some important points
as well as speculating about the possibility of use the $\Theta$
operator in order to map results from Dirac lagrangian into ELKO
lagrangian and vice-versa.

\section{ELKO spinor fields}

In this Section  the formal properties of ELKO spinor fields  are
briefly revised \cite{allu,alu2,alu3}. An ELKO, denoted by $\Psi$,
corresponding to a plane wave with momentum $p=(p^{0},\mathbf{p)}$
can be written, without loss of generality, as
$\Psi(p)=\lambda({\bf p}) e^{-i{p\cdot x}}$ (or
$\Psi(p)=\lambda({\bf p}) e^{i{p\cdot x}}$) where
\begin{equation}
\lambda({\bf
p})=\binom{i\Phi\phi_{L}^{\ast}(\mathbf{p})}{\phi_{L}(\mathbf{p})},
\label{1}%
\end{equation}
\noindent  $\phi_{L}(\mathbf{p})$ denotes a left-handed Weyl
spinor, and given the rotation generators denoted by
${\mathfrak{J}}$, the Wigner's spin-1/2 time reversal operator
$\Phi$ satisfies $\Phi
\mathfrak{J}\Phi^{-1}=-\mathfrak{J}^{\ast}$. Hereon, as in
\cite{allu}, the Weyl representation of $\gamma^{\mu}$ is used,
i.e.,
\begin{equation}
\gamma_{0}=\gamma^{0}=%
\begin{pmatrix}
\OO & \II\\
\II & \OO
\end{pmatrix}
,\quad-\gamma_{k}=\gamma^{k}=%
\begin{pmatrix}
\OO & -\sigma_{k}\\
\sigma_{k} & \OO
\end{pmatrix}
,\quad
\gamma^{5}=-i\gamma^{0}\gamma^{1}\gamma^{2}\gamma^{3}=-i\gamma^{0123}=
\begin{pmatrix}
\II&\OO \\
\OO & -\II
\end{pmatrix}
\label{dirac matrices}%
\end{equation}
\noindent where
\begin{equation}
\II= \begin{pmatrix}
1 & 0\\
0 & 1
\end{pmatrix}
,\quad \OO=\begin{pmatrix}
0 & 0\\
0 & 0
\end{pmatrix}
,\quad
\sigma_{1}=%
\begin{pmatrix}
0 & 1\\
1 & 0
\end{pmatrix}
,\quad\sigma_{2}=%
\begin{pmatrix}
0 & -i\\
i & 0
\end{pmatrix}
,\quad\sigma_{3}=%
\begin{pmatrix}
1 & 0\\
0 & -1
\end{pmatrix}.
\end{equation}
\noindent    ELKO spinor fields are eigenspinors of the charge
conjugation operator $C$, i.e., $C\lambda(\bf{p})=\pm \lambda({\bf
p})$, for $
C=%
\begin{pmatrix}
\OO & i\Phi \\
-i\Phi & \OO
\end{pmatrix}\,K.$  The operator $K$ is responsible for the $\mathbb{C}$-conjugation of
spinor fields appearing on the right. The plus sign stands for
{\it self-conjugate} spinors, $\lambda^{S}({\bf p})$, while the
minus yields {\it anti self-conjugate} spinors, $\lambda^{A}({\bf
p})$. Explicitly, the complete form of ELKO spinor fields can be
found by solving the equation of helicity
$(\sigma\cdot\widehat{\bf{p}})\phi^{\pm}(\mathbf{0})=\pm
\phi^{\pm}(\mathbf{0})$ in the rest frame and subsequently
performing a boost, in order  to recover the result for any ${\bf
p}$ \cite{allu}. Note that the helicity of
$i\Phi[\phi_{L}(\mathbf{p})]^\ast$ is opposed to that of
$\phi_L(\mathbf{p})$, since
$(\sigma\cdot\widehat{\bf{p}})\Phi[\phi_L^{{\pm}}(\mathbf{0})]^\ast=\mp
\Phi[\phi_L^{\ast\pm}(\mathbf{0})]^\ast$. Here
$\widehat{\bf{p}}:={\bf p}/\|{\bf
p}\|=(\sin\theta\cos\phi,\sin\theta\sin\phi,\cos\theta)$. The four
spinor fields are given by
\begin{equation}
\lambda^{S/A}_{\{\mp,\pm \}}({\mathbf
p})=\sqrt{\frac{E+m}{2m}}\Bigg(1\mp \frac{{\bf
p}}{E+m}\Bigg)\lambda^{S/A}_{\{\mp,\pm \}}(\bf{0}),
\label{form}\end{equation}
\begin{equation} \qquad\text{where}\qquad
\lambda_{\{\mp,\pm \}}(\bf{0})=%
\begin{pmatrix}
\pm i \Theta[\phi^{\pm}(\bf{0})]^{*} \\
\phi^{\pm}(\bf{0})
\end{pmatrix}
\label{four}.\end{equation} The phases are adopted so that
    \begin{eqnarray}
    \phi^+({\mathbf{0}})= \sqrt{m}\left(\begin{array}{c}
    \cos(\theta/2) e^{-i \phi/2}\\
    \sin(\theta/2) e^{i \phi/2}
    \end{array}\right),\qquad
    \phi^-({\mathbf{0}}) = \sqrt{m}\left(\begin{array}{c}
    -\sin(\theta/2) e^{-i \phi/2}\\
    \cos(\theta/2) e^{i \phi/2}
    \end{array}\right), \label{2222}
    \end{eqnarray}\noi at rest, and since
$\Theta[\phi^{\pm}(\bf{0})]^{*}$ and $\phi^{\pm}(\bf{0})$ present
opposite helicities, ELKO cannot be an eigenspinor field of the
helicity operator, and indeed carries both helicities. In order to
guarantee an invariant real norm, as well as positive definite
norm for two ELKO spinor fields, and negative definite norm for
the other two, the ELKO dual is given by \cite{allu}
\begin{equation}
\overset{\neg}{\lambda}^{S/A}_{\{\mp,\pm \}}({\bf p})=\pm i \Big[
\lambda^{S/A}_{\{\pm,\mp \}}({\bf p})\Big]^{\dag}\gamma^{0}
\label{dual}.\end{equation}\noindent
 It is useful to
choose $i\Theta=\sigma_{2}$, as in \cite{allu}, in such a way that
it is possible to express
\begin{equation}
\lambda(\mathbf{p})=\binom{\sigma_{2}\phi_{L}^{\ast}(\mathbf{p})}{\phi_{L}(\mathbf{p})}.
\label{01}%
\end{equation}

Now, any flagpole spinor field is an eigenspinor field of the
charge conjugation operator \cite{lou1,lou2}, here represented by
$\mathcal{C}\psi=-\gamma ^{2}\psi^{\ast}$. Indeed
\begin{align}
-\gamma^{2}\lambda^{\ast}  &  =%
\begin{pmatrix}
0 & \sigma_{2}\\
-\sigma_{2} & 0
\end{pmatrix}
\binom{(\sigma_{2}\phi^{\ast})^{\ast}}{\phi^{\ast}}
=\binom{\sigma_{2}\phi^{\ast}}{-\sigma_{2}\sigma_{2}^{\ast}\phi}\nonumber
=\lambda.
\end{align}
\noindent

\section{Mapping between Dirac and ELKO spinor fields}
In this Section we briefly review which are the Dirac spinor
fields that can be led to ELKO spinor fields. First introduce a
matrix $M\in\CC(4)$ that defines the transformation from an
\emph{a priori} arbitrary Dirac spinor field to an ELKO spinor
field, i.e., \bege\label{cond} M\psi = \lambda.\enge\noi In
\cite{osmano} it was proved that not all DSFs can be led to ELKO,
but only a subset of the three classes --- under Lounesto
classification --- of DSFs restricted to some conditions.
Explicitly writing the entries of $M = [m_{pq}]_{p,q=1}^4$,  we
showed in \cite{osmano} that a general form for $M$ is given by
\begin{equation}
M=%
\begin{pmatrix}
m_{11} & m_{12} & -\chi m_{11} & -i\epsilon \ka -\chi m_{12}\\
m_{21} & m_{22} & i\epsilon \ka-\chi m_{21} & -\chi m_{22}\\
m_{31} & m_{32} & 1-\chi m_{31} & -\chi m_{32}\\
m_{41} & m_{42} & -\chi m_{41} & 1- \chi m_{42}\\
\end{pmatrix}
\label{eme2},\end{equation} where $\phi_{R}({\bf p})=\chi
\phi_{L}({\bf p})$, and $\chi = \frac{E + {\mathbf{\sigma}}\cdot
{\mathbf{p}}}{m}$ and $\ka\psi = \psi^*$. It should be emphasized
that $M$ is not unique, and also, in \cite{osmano} we completely
fixed the matrix $M$, inserting the {\it ansatz}
\begin{eqnarray}
m_{11}=m_{22}=0= m_{32}=m_{41},\nonumber\\
m_{31}=m_{42}=1=m_{12} \label{fix}
\end{eqnarray} in such a way that  $M$ is written as
\begin{equation}
M=%
\begin{pmatrix}
0 & 1 & 0 & -i\epsilon \ka -\chi \\
-1 & 0 & i\epsilon \ka+\chi & 0\\
1 & 0 & 1-\chi & 0\\
0 & 1 & 0 & 1- \chi \\
\end{pmatrix}
\label{ansatz}.\end{equation} \noindent Note that such matrix is
not unitary, and since $\det M\neq 0$, there exists (see
Eq.(\ref{cond})) $M^{-1}$ such that $\psi=M^{-1}\lambda$. Besides,
it is immediate to note that \bege
\bar{\psi}:=\psi^\dagger\gamma^0=\lambda^{\dag}(M^{-1})^{\dag}\gamma^{0},\label{j1}
\enge such that $\bar\psi$ can be related to the ELKO dual by
\bege\bar{\psi} =\mp
i\overset{\neg}{\lambda}^{S/A}_{\{\mp,\pm\}}\gamma^{0}(M^{-1})^{\dag}\gamma^{0}.\label{j2}
\enge In what follows, the matrix $M$ establishes necessary
conditions on the Dirac spinor fields under which the mapping
given by Eq.(\ref{cond}) is satisfied. However, the {\it ansatz}
in Eq.(\ref{ansatz}) has just an illustrative r\^ole. In fact, for
any matrix satisfying Eq.(\ref{eme2}), there are corresponding
constraints on the components of DSFs. Hereafter, we shall
calculate the conditions to the case where {\bf p} = 0 (and
consequently $\chi = 1$), since a Lorentz boost can be implemented
on the rest frame in the constraints. Anyway, without lost of
generality, the conditions to be found on DSFs must hold in all
referentials, and in particular in the rest frame corresponding to
{\bf p} = 0.

Denoting the Dirac spinor field as $\psi =
(\psi_1,\psi_2,\psi_3,\psi_4)^T$ ($\psi_r \in \CC, r=1,\ldots,4$),
 we achieved simultaneous conditions for
the Dirac spinor field be led to an ELKO spinor field
respectively:
\begin{align}
0&= \mathbb{R}{\rm e}(\psi_1^*\psi_3)+\mathbb{R}{\rm
e}(\psi_2^*\psi_4)\n 0&= \mathbb{R}{\rm e} (\psi_2^*\psi_3)+
\mathbb{R}{\rm e} (\psi_1^*\psi_4)\n 0&=
\mathrm{Im}(\psi_1^*\psi_4)-\mathrm{Im}(\psi_2^*\psi_3)-2\mathrm{Im}(\psi_3^*\psi_4)-2\mathrm{Im}(\psi_1^*\psi_2)
\n 0&= \mathbb{R}{\rm e}(\psi_1^*\psi_3)-\mathbb{R}{\rm
e}(\psi_2^*\psi_4) \label{partes}.\end{align}  Note that the first
and the last conditions together mean $\mathbb{R}{\rm
e}(\psi_1^*\psi_3)=0$ and $\mathbb{R}{\rm e}(\psi_2^*\psi_4)=0$.
In what follows we obtain the extra necessary and sufficient
conditions for each class of Dirac spinor fields, according to
Lounesto spinor fields classification (See Appendix).

Using the decomposition $\psi_j=\psi_{ja}+i\psi_{jb}$ (where
$\psi_{ja}$ = $\mathbb{R}$e($\psi_j$) and $\psi_{jb}$ =
Im($\psi_j$)) it follows that $\mathbb{R}{\rm
e}(\psi_i^*\psi_j)=\psi_{ia}\psi_{ja}+\psi_{ib}\psi_{jb}$ and
$\mathrm{Im}(\psi_i^*\psi_j)=\psi_{ia}\psi_{jb}-\psi_{ib}\psi_{ja}$
for $i,j=1,\ldots,4$. So, in components, the conditions in common
for all types of DSFs are \begin{eqnarray}
\psi_{1a}\psi_{3a}+\psi_{1b}\psi_{3b}&=&0 \label{c1},\\
\psi_{2a}\psi_{4a}+\psi_{2b}\psi_{4b}&=&0 \label{c2},
\end{eqnarray} and the additional conditions for each case are
summarized in Table I below.
\begin{center}
\begin{table}[!h]
\begin{tabular}{|c|c|}
  \hline
  {\bf Class} & {\bf Additional conditions}  \\
  \hline
 \hline (1) & $\psi_{2a}(\psi_{3a}-\psi_{3b})+\psi_{2b}(\psi_{3a}+\psi_{3b}) = 0 = \psi_{3a}\psi_{4b}-\psi_{3b}\psi_{4a}$ \\\hline
  (2) & $\psi_{3a}\psi_{4b}-\psi_{3b}\psi_{4a} = 0 = \psi_{2a}\psi_{3a}+\psi_{2b}\psi_{3b}+\psi_{1a}\psi_{4a}+\psi_{1b}\psi_{4b}$
\\\hline
  (3) & $\psi_{2a}(\psi_{3a}-\psi_{3b})+\psi_{2b}(\psi_{3a}+\psi_{3b})=0$
 and \\
{}&$(\psi_{1a}\psi_{4b}-\psi_{1b}\psi_{4a})-(\psi_{2a}\psi_{3b}-\psi_{2b}\psi_{3a})-2(\psi_{3a}\psi_{4b}-\psi_{3b}\psi_{4a})-$
 $2(\psi_{1a}\psi_{2b}-\psi_{1b}\psi_{2a})=0$ \\
  \hline
\end{tabular}
\caption{Additional conditions, in components, for class (1), (2)
and (3) Dirac spinor fields.}
\end{table}
\end{center}

Once the matrix $M$ --- leading an arbitrary Dirac spinor field to
an ELKO ---  has been introduced, we proved that it can be written
 in the general form given by Eq.(\ref{eme2}), without loss of generality. The \emph{ansatz} given by Eq.(\ref{eme2})
is useful to illustrate and explicitly exhibit how to obtain the
necessary conditions on the components of a DSF --- under Lounesto
spinor field classification (see Appendix, for more details and
references therein --- in order to it be led to an ELKO spinor
field. In the case of a type-(1) Dirac spinor field there are six
conditions, and then the equivalence class of type-(1) Dirac
spinor fields
 that can be led to ELKO spinor fields  can be written in the form\footnote{Among the
 three equivalent definitions of spinor fields, viz., the classical, algebraic, and operatorial, here
the classical one --- where a spinor is an element that carries
the representation space of the group Spin$_+$(1,3),
 is regarded.}
\bege \psi=\begin{pmatrix}
\psi_1\\
f_1(\psi_1)\\
f_2(\psi_1)\\
f_3(\psi_1)\\
\end{pmatrix}
\enge\noi where $f_i$ are complex scalar functions of the
component $\psi_1\in\mathbb{C}$ of $\psi$, obtainable --- using
the Implicit Function Theorem --- through the conditions given in
Eqs.(\ref{c1}), (\ref{c2}), and also those given by Table I. For a
general and arbitrary \emph{ansatz}, the equivalence class of
type-(1) DSFs
 that can be led to ELKO spinor fields, via the matrix $M$, are given by
\bege\psi=\begin{pmatrix}
\psi_1\\
g_1(M)(\psi_1)\\
g_2(M)(\psi_1)\\
g_3(M)(\psi_1)\\
\end{pmatrix}
\enge\noi where each $g_i(M)$ is a complex scalar function of the
component $\psi_1\in\mathbb{C}$ of $\psi$. Such scalar functions
depend explicitly on the form of $M$, and to a fixed but arbitrary
$M$ there corresponds other six conditions analogous to
Eqs.(\ref{c1}), (\ref{c2}), and also those given by Table I. All
these conditions obtained by the \emph{ansatz} is general, and
illustrates the general procedure of finding the conditions. For
the equivalence class of type-(2) and -(3) DSFs that are led to
ELKO spinor fields, it is only demanded five conditions, instead
of six \cite{osmano}. In both cases, the most general form of the
DSFs are given by \bege\psi=
\begin{pmatrix}
\psi_{1a} + i\psi_{1b}\\
\psi_{2a} + i\psi_{2b}\\
\psi_{3a} + i\psi_{3b}\\
\psi_{4a} + i\psi_{4b}\\
\end{pmatrix}=\begin{pmatrix}
\psi_{1a} + i\psi_{1b}\\
\psi_{2a} + i h_1(M)(\psi_{1a},\psi_{1b},\psi_{2a})\\
h_2(M)(\psi_{1a},\psi_{1b},\psi_{2a}) + i h_3(M)(\psi_{1a},\psi_{1b},\psi_{2a})\\
h_4(M)(\psi_{1a},\psi_{1b},\psi_{2a}) + i h_5(M)(\psi_{1a},\psi_{1b},\psi_{2a})\\
\end{pmatrix}
\enge\noi  where each $h_A(M)$ ($A=1,\ldots,5$) is a $M$
matrix-dependent real scalar function of the (real) components
$\psi_{1a},\psi_{1b},\psi_{2a}$
 of $\psi$. For more details and considerations see \cite{osmano}.

 To summarize,
 the development reviewed in this Section is accomplished without any loss of generality.
For more details and considerations see \cite{osmano}. In the next
Section we shall to introduce an operator, constructed upon $M$,
responsible to lead the Dirac action into the ELKO action, apart
from a surface term.

\section{Mapping the Dirac Lagrangian in the ELKO Lagrangian}

In this Section  the action of the mass dimension-transmuting
$\Theta$ operator on the Dirac Lagrangian is explicitly derived,
leading to the ELKO Lagrangian plus a surface term. When the
transformation of  the  spinor fields constituting the Lagrangian
is dealt, there are two possibilities to construct the mass
dimension-transmuting $\Theta$ operator action: the fundamental
$(\Theta \LLL_{\rm Dirac})$ and the adjoint action $(\Theta
\LLL_{\rm Dirac}\Theta^{-1})$. From a formal viewpoint, it is
always possible to define the adjoint action $(\Theta \LLL_{\rm
Dirac}\Theta^{-1})$, resulting in \beq
\Theta \LLL_{\rm Dirac}\Theta^{-1}&=&\tth[(\bar\psi\g^\mu\pa_\mu\psi)-m_D(\bar\psi\psi)]\tth^{-1}\nonumber\\
&=&\tth\bar\psi\tth^{-1}\tth\g^\mu\tth^{-1}\tth\pa_\mu\psi\tth^{-1}-m_D\tth\bar\psi\tth^{-1}\tth\psi\tth^{-1}.
\eeq

There is no reason \emph{a priori} for fixing one of the two
actions, although the adjoint action resembles some correspondent
physical symmetry, in the case it is performed by an unitary
operator. But it can be shown that there \emph{is not} an unitary
operator able to lead the Dirac Lagrangian to the ELKO Lagrangian,
and the Dirac and ELKO spinor fields present respectively local
and non-local properties. It follows immediately from the results
in \cite{allu,osmano}.

If $\tth$ does exist as a true physical operator, the extended
Lagrangian density describes  the SM in order to incorporate dark
matter fields \cite{gau}. Hereon the fundamental action is taken
into account, by defining an (even) derivation $\tth$ on the space
of spinor fields ($\tth(\psi\phi)=(\tth\psi)\phi +
\psi(\tth\phi)$, where $\psi,\phi$ denote spinor fields) such that
$\tth i = i\tth$. Such an operator acts on $\LLL_{\rm Dirac}$ as
\beq \tth\LLL_{\rm Dirac}&=&\tth(i\bar\psi\g^\mu\pa_\mu\psi-m_D\bar\psi\psi)\nonumber\\
&=&i\tth(\bar\psi\g^\mu\pa_\mu\psi)-m_D\tth(\bar\psi\psi)\nonumber\\
&=&i[(\tth\bar\psi)\g^\mu\pa_\mu\psi+\bar\psi\tth(\g^\mu\pa_\mu\psi)]-m_D[\tth(\bar\psi)\psi
+ \bar\psi(\tth\psi)].\label{colocada}\eeq By straightforward
algebraic manipulation, this equation can be expressed as
\beq\label{dir} \tth\LLL_{\rm
Dirac}&=&(\tth\bar\psi)(i\slash\hspace{-1.6mm}\pa-m_D)\psi
+i\bar\psi(\tth\g^\mu)\pa_\mu\psi+
i\bar\psi\g^\mu\pa_\mu(\tth\psi)-i\bar\psi\g^\mu(\pa_\mu\tth)\psi -m_D\bar\psi(\tth\psi)\nonumber\\
&=&(\tth\bar\psi)(i\slash\hspace{-1.6mm}\pa-m_D)\psi
+\bar\psi(i\slash\hspace{-1.6mm}\pa-m_D)(\tth\psi)+
i\bar\psi(\tth\g^\mu)\pa_\mu\psi-i\bar\psi\g^\mu(\pa_\mu\tth)\psi,\eeq
where $\slash\hspace{-1.6mm}\pa=\g^\mu\pa_\mu$.

Now, we construct the spinor space derivation $\tth$ satisfying
the required property $\tth i = i\tth$ in such a way that its
action on $\bar\psi$ is given by \beq\label{2}
\tth\bar\psi=-\frac{m^2}{2}\bar\psi\g^0M^\dagger i\g^0M
(i\slash\hspace{-1.8mm}\pa-m_D)^{-1}, \eeq \noi where $M$ is the
matrix introduced in the previous Section. Note that the
nontrivial form of Eq.(\ref{2}) is motivated and justified by the
property \beq \label{eq}
(\tth\bar\psi)(i\slash\hspace{-1.6mm}\pa-m_D)\psi&=& -\frac{m^2}{2}\bar\psi\g^0M^\dagger i\g^0M \psi\nonumber\\
&=&-\frac{m^2}{2}\lam^\dagger(M^{-1})^\dagger\g^0\g^0M^\dagger
i\g^0 MM^{-1}\lam
\nonumber\\
&=& -\frac{m^2}{2}\du ,\eeq \noi which is exactly half the mass
term in ELKO Lagrangian \cite{allu,alu2,alu3}. The mass dimension
associated with the $m_{D}$ parameter equals the mass dimension of
$m$. We shall, however, preserve this notation in order to avoid
confusion. In Eq.(\ref{eq}) we used the fact that $M\psi=\lam$ and
 $\psi=M^{-1}\lam$, where $\psi$ is a Dirac spinor field and
$\lam$ denotes an ELKO \cite{osmano}. Also $\bar\psi=\lam^\dagger
(M^{-1})^\dagger\g^0$. In addition, it is possible to apply the
operator $\tth$, satisfying the same required property
 $\tth i = i\tth$, on the Dirac dual spinor field in such a way that
\beq\label{3}
\tth\psi=-(i\slash\hspace{-1.6mm}\pa-m_D)^{-1}\frac{m^2}{2}\g^0M^\dagger
i\g^0M\psi \eeq \noi and, in a similar reasoning related to the
derivation of Eq.(\ref{eq}), the mass term in ELKO Lagrangian is
given by
$\bar\psi(i\slash\hspace{-1.6mm}\pa-m_D)\tth\psi=-\frac{m^2}{2}\du$.
The action of the symmetry operator $\tth$ on the Dirac Lagrangian
can be expressed, using Eq.(\ref{dir}), as \beq\tth\LLL_{\rm
Dirac} &=&
i\bar\psi[(\tth\g^\mu)\pa_\mu-\g^\mu(\pa_\mu\tth)]\psi-m^2\du.\eeq

The $\Theta$ operator always exists, in the sense that it is
constructed in terms of the Dirac operator derivatives, of the
Dirac matrices, and also in terms of $M$. As these three
objects always exist, $\Theta$ also always exists. Since $M$ is
not unique --- it is given by Eq.(\ref{eme2}) --- so the operator
$\Theta$ is also \emph{not} unique. But once we fix the operator
$M$, as in Eq.(\ref{ansatz}), the operator $\Theta$ is
automatically unique.

On the other hand, the general form of the ELKO
Lagrangian\footnote{Without taking into account auto-interaction
terms \cite{alu2}.} is given by the expression \cite{alu2}
\beq\label{elkol} \LLL_{\rm ELKO}=\pa_\mu\dual{\lam}\pa^\mu\lam -
m^2\dual{\lam}\lam . \eeq \noi Since the transformed Dirac
Lagrangian presents the same mass term as the ELKO Lagrangian
\cite{alu2}, it is possible to relate the two Lagrangians, using
the relations between ELKO and Dirac spinors fields, as
\beq\label{inserida}\tth\LLL_{\rm Dirac} &=&
i\bar\psi[(\tth\g^\mu)\pa_\mu-\g^\mu(\pa_\mu\tth)]\psi+\LLL_{\rm
ELKO}-\pa_\mu\dual{\lam} \pa^\mu\lam .\eeq Integrating by parts,
and using the relations $\psi=M^{-1}\lam$, $\bar\psi=\lam^\dagger
(M^{-1})^\dagger\g^0$, and Eq.(\ref{dual}), it reads \beq
\label{a1}\tth\LLL_{\rm Dirac} &=&\LLL_{\rm ELKO}
-\pa_\mu(\dual{\lam}\pa^\mu\lam)
+\dual\lam\Big[\left(\g^0(M^{-1})^\dagger
\g^0[(\tth\g^\mu)\pa_\mu-\g^\mu(\pa_\mu\tth)]M^{-1}\right)
+\Box\Big]\lam . \eeq \noindent This remarkable expression shows,
in the scope illustrated above, a strong relationship between
$\tth\LLL_{\rm Dirac}$ and $\LLL_{\rm ELKO}$. The second term in
Eq.(\ref{a1}) is a surface term and can be neglected by
appropriate boundary conditions. Consequently, it immediately
follows from Eq.(\ref{a1}) that  \beq\label{a2}\hspace*{-6mm} \int
\tth\LLL_{\rm Dirac}\,d^4x &=& \int\LLL_{\rm ELKO}\,d^4x +\int
\dual\lam\Big[\left(\g^0(M^{-1})^\dagger
\g^0[(\tth\g^\mu)\pa_\mu-\g^\mu(\pa_\mu\tth)]M^{-1}\right)+\Box\Big]\lam\,d^4x.
\eeq

Let us define the $\mathcal{P}$ operator as \beq
\label{a3}\mathcal{P}= \g^0(M^{-1})^\dagger
\g^0(\slash\hspace{-1.6mm}\pa\cdot\tth)M^{-1}+\Box,\eeq where
$\slash\hspace{-1.6mm}\pa\cdot\tth =
(\Theta\gamma^{\mu})\pa_{\mu}-\gamma^{\mu}(\pa_{\mu}\Theta)$.
Here, given the spacetime metric $g$, the left contraction is
implicitly defined by $g(a\cdot b,c)=g(b,\tilde{a}\wedge c)$,
where $a,b,c$ denote elements of the Dirac algebra and $\tilde{a}$
denotes the reversion anti-automorphism on
 the Dirac algebra element $a$ \cite{osmano,lou1}. In addition, $a\wedge b$ denotes the exterior product
 between two elements $a, b$ in the Dirac algebra. Now, Eq.(\ref{a2}) is suitable written as \beq\label{a4}\int
\tth\LLL_{\rm Dirac}\;d^4x &=& \int\LLL_{\rm ELKO}\;d^4x +\int
\dual\lam\mathcal{P}\lam\; d^4x. \eeq

In order to define the action $\tth$ that is responsible to map
$\LLL_{\rm Dirac}$ to $\LLL_{\rm ELKO}$, we must restrict the
action $\tth$ on the spinor space, eliminating then the extra term
given by $\int \dual\lam\mathcal{P}\lam\; d^4x$ in Eq.(\ref{a4}).
From Eq.(\ref{a3}), by imposing $\mathcal{P}\equiv 0$, it follows
that \beq\label{a6} \g^0(M^{-1})^\dagger
\g^0(\slash\hspace{-1.6mm}\pa\cdot\tth)M^{-1}=-\Box , \eeq \noi
which can be expressed by \beq\label{a7}
(\slash\hspace{-1.6mm}\pa\cdot\tth) &=& -\g^0 M^\dagger
\g^0 \Box M\nonumber\\
&=& (\g^0 M^\dagger i\g^0 M)M^{-1} i \Box M \eeq and, also, using
the most general form of the matrix $M$ that leads a Dirac spinor
field to an ELKO and presents the properties
 $Mi=iM^\ast,\;iM=M^\ast i$ \cite{osmano}, Eq.(\ref{a7}) reads
\beq\label{a10} (\slash\hspace{-1.6mm}\pa\cdot\tth) &=&  (\g^0
M^\dagger i\g^0 M)M^{-1}  \Box M^\ast i . \eeq The equation above
is an additional formal constraint which must be respected by the
$\tth$ operator in order to eliminate the contribution from the
$\mathcal{P}$ operator. In order to relate this constraint to the
action of $\tth$ in $\psi$ and $\bar\psi$, note that Eqs.(\ref{2})
and (\ref{3}) can be respectively written as \beq\label{12}
(\tth\bar\psi)(i\slash\hspace{-1.8mm}\pa-m_D)=-\frac{m^2}{2}\bar\psi(\g^0M^\dagger
i\g^0M)
 \eeq \noi
\noi and \beq\label{13}
(i\slash\hspace{-1.8mm}\pa-m_D)(\tth\psi)=-\frac{m^2}{2}(\g^0M^\dagger
i\g^0M)\psi . \eeq \noi Now, using Eq.(\ref{a10}), the action of
the operator $\tth$ on the spinor space arises from Eqs.(\ref{12})
and (\ref{13}), and completely defines the action of the symmetry
$\Theta$ in the Dirac Lagrangian, in order to lead it to the ELKO
Lagrangian. It is given explicitly by \beq\label{16}
\tth\bar\psi=i\frac{m^2}{2}\bar\psi[(\slash\hspace{-1.6mm}\pa\cdot\tth)
(\Box M^\ast)^{-1} M]\,({i\slash\hspace{-1.6mm}\pa-m_D})^{-1}
 \eeq
\noi and \beq\label{17}
\tth\psi=i\frac{m^2}{2}({i\slash\hspace{-1.6mm}\pa-m_D})^{-1}\bar\psi\
[(\slash\hspace{-1.6mm}\pa\cdot\tth)(\Box M^\ast)^{-1} M]\psi
.\eeq \noi We emphasize that the
$(\slash\hspace{-1.6mm}\pa\cdot\tth)$ term appearing in the
formulae above is fixed by Eq.(\ref{a10}) in terms of the matrix
$M$ that turns Dirac to ELKO spinor fields \cite{osmano}.

After all, from Eq.(\ref{a2}) it is easy to see that
\beq\label{a111}\int(\tth\LLL_{\rm Dirac})\;d^4x &=&\int\LLL_{\rm
ELKO}\;d^4x .\eeq \noi In this form, the operator $\tth$ (acting
according to (\ref{16}), (\ref{17}) and (\ref{a10})) is
responsible to lead the Dirac spinor field lagrangian to the ELKO
lagrangian, which suggests the possibility of extending the
Standard Model of elementary particles in order to incorporate
dark matter. We shall to make some comments about this results in
the next Section.

\section{Concluding Remarks and Outlooks}

We have constructed an operator $\tth$ intended to
transmute the spinor fields which have mass dimension  1 to spinor fields presenting mass dimension 3/2, and
vice-versa. It is obtained via the appropriate definition of the
$\tth$ operator acting as an even derivation on the spinor space.
This action of the $\tth$ operator upon the Dirac Lagrangian is
intended to transform all the dynamics of a free Dirac spinor
field into an also free ELKO spinor field in the classical field
theoretical framework. Besides, we have shown that in order to the
Dirac Lagrangian to be suitably led to the ELKO Lagrangian,  the
action of the $\tth$ operator, defined on both spinors $\psi$ and
conjugate spinor $\bar\psi$, must obey an additional constraint
given by Eq.(\ref{a10}). The actions Eqs.(\ref{16}) and (\ref{17})
constitute a minimal set that completely defines the operator $\tth$,
and although it seems $\tth\psi$ can be written in terms of
$\tth\bar\psi$, as the unique difference between them is the
common terms $[(\slash\hspace{-1.6mm}\pa\cdot\tth) (\Box
M^\ast)^{-1} M]$ and $({i\slash\hspace{-1.6mm}\pa-m_D})^{-1}$
appearing in reverse order, each of these two terms does not
possess a definite homogeneous multivectorial structure in the
spacetime Dirac algebra. Consequently, if we attempt to express,
e.g.,  $\tth\psi$ in terms of $\tth\bar\psi$, both the terms
$[(\slash\hspace{-1.6mm}\pa\cdot\tth) (\Box M^\ast)^{-1} M]$ and
$({i\slash\hspace{-1.6mm}\pa-m_D})^{-1}$ are modified, when the
anti-automorphism that reverses the multivectorial structure in
the spacetime Dirac algebra acts on their product. The reversion
changes the order of these terms, and in addition modifies each
one of the terms $[(\slash\hspace{-1.6mm}\pa\cdot\tth) (\Box
M^\ast)^{-1} M]$ and $({i\slash\hspace{-1.6mm}\pa-m_D})^{-1}$.

We should point out a few more remarks about the future of this
line of research. First, in the scope of the formalism developed
in this paper, the constraint (\ref{a10}) is just a pure formal
restriction, and does not have an immediate physical
interpretation yet. A deeper investigation on such a constraint
can reveal some important physical implication about a possible
extension of the Standard Model. Second, the transformation
analyzed here was obtained in the scope of classical field theory.
In quantum field language, ELKO spinor fields are extended
objects, while Dirac spinor fields belongs to a standard Wigner's
class \cite{allu,alu2}. The quantum version of the $\tth$ symmetry
should stress this fact. From the functional form of $\tth$ acting
in $\psi$ and $\bar\psi$ we see the presence of an object that, in quantum
field theory, is identified to the Dirac propagator. This may suggest that the
non-local character of the ELKO spinor field is already taken into
account. However, obviously, the transformation of a local quantum
object into a non-local one brings a deep modification of the
canonical structure of the theory. Regarding the formalism
developed here, we prefer to be more conservative and not to
relate the form of the $\tth$ operator to any ``quantum
adjective''. 

We should emphasize however that, even in the scope
of classical field theory, the main results of this paper can be
used in order to map solutions from the Dirac lagrangian to ELKO
lagrangian and vice-versa, respecting the conditions under which
the actions of both systems are related. It seems that, again, the
formal constraint (\ref{a10}) plays a crucial role in such a map.
Finally, the study of the solutions that can be led from one system
to another seems to provide an useful field of research, since the
ELKO spinor field is, in many aspects, more tractable than the
Dirac spinor field. On the other hand, in taking this program
seriously, the vast literature concerning DSF's can bring a lot of
physical information about ELKO spinor field itself.

As a last remark, the formalism developed in Section IV can be
immediately applied to the dark matter candidate described in
\cite{gau}. There are proposed new fields describing dark matter
and suggesting how  the dark matter sector Lagrangian density
arises from the mass dimension-transmuting symmetry $\tth$, here
investigated, constructed upon the algebraic mapping of Dirac
spinor fields into ELKO spinor fields. It is one more contribution
in order to extend the Standard Model to incorporate dark matter.

In particular, there has a lot of advances accomplished by the refs. \cite{allu,alu2,alu3,boe1}
where there has been explored the relationship between a quantum field theory related to ELKO spinor
fields and some applications in an attempt to describe and investigate dark matter. Once a prescription
between Dirac theory and ELKO is accomplished, it could be easier and useful to investigate some properties 
of dark matter formalism, using the Dirac formalism for relativistic quantum mechanics and quantum field theory.
In addition, using the mapping between ELKO spinor fields and the three classes Dirac spinor fields, it was
shown in \cite{nossoijg} that the Einstein-Hilbert, the Einstein-Palatini, and the
Holst actions can be derived from the Quadratic Spinor Lagrangian --- used  as the prime Lagrangian for supergravity. The Holst action is related to
the Ashtekar's quantum gravity formulation, and shows up also in the proof of
gravitational theory as a SUSY gauge theory as well \cite{tung2}. To each one of
these classes, there corresponds a unique kind of action for a
covariant gravity theory \cite{ijmpdze}.

There is a lot of developments based on the present formalism of ELKO action and the Dirac action
that can bring some new light on some related research lines.
\section{Acknowledgment}
The authors are very grateful to Prof. Dharamvir Ahluwalia for
important comments about this paper, and to the IJMPA Referee as
well. Rold\~ao da Rocha thanks to Funda\c c\~ao de Amparo \`a
Pesquisa do Estado de S\~ao Paulo (FAPESP) (2008/06483-5) and J.
M. Hoff da Silva thanks to CAPES-Brazil for financial support.

\section*{Appendix: Bilinear Covariants, Lounesto Spinor Field Classification  and ELKO spinor fields}

This Appendix is devoted to recall the bilinear covariants, using
the programme introduced in \cite{wal}, just for book keeping
purposes. In this article all spinor fields live in Minkowski
spacetime $(M,\eta ,D,\tau_{\eta},\uparrow)$. The manifold $M$
$\simeq\mathbb{R}^{4}$, $\eta$ denotes a constant metric, where
$\eta(\partial/\partial x^{\mu},\partial/\partial x^{\nu
})=\eta_{\mu\nu}=\mathrm{diag}(1,-1,-1,-1)$, $D$ denotes the
Levi-Civita connection associated with $\eta$, $M$ is oriented by
the 4-volume element $\tau_{\eta}$ and time-oriented by
$\uparrow$.  Here $\{x^{\mu}\}$ denotes global coordinates in the
Einstein-Lorentz gauge, naturally
adapted to an inertial reference frame  $\mathbf{e}_{0}%
=\partial/\partial x^{0}$. Let $\mathbf{e}_{i}=\partial/\partial
x^{i}$, $i=1,2,3$.
 Also,
$\{\mathbf{e}_{\mu}\}$ is a section of the frame bundle
$\mathbf{P}_{\mathrm{SO}_{1,3}^{e}}(M)$ and
$\{\mathbf{e}^{\mu}\}$ is its reciprocal frame satisfying $\eta(\mathbf{e}%
^{\mu},\mathbf{e}_{\nu}):=\mathbf{e}^{\mu}\cdot\mathbf{e}_{\nu}=\delta_{\nu
}^{\mu}$. Classical spinor fields carrying a $D^{(1/2,0)}\oplus
D^{(0,1/2)}$, or $D^{(1/2,0)}$, or $D^{(0,1/2)}$ representation of
SL$(2,\mathbb{C)\simeq }\;\,\mathrm{Spin}_{1,3}^{e}$ are sections
of the vector bundle $
\mathbf{P}_{\mathrm{Spin}_{1,3}^{e}}(M)\times_{\rho}\mathbb{C}^{4},
$ where $\rho$ stands for the $D^{(1/2,0)}\oplus D^{(0,1/2)}$ (or
$D^{(1/2,0)}$
or $D^{(0,1/2)}$) representation of SL$(2,\mathbb{C)\simeq}\;\,\mathrm{Spin}%
_{1,3}^{e}$ in $\mathbb{C}^{4}$.  Given a spinor field $\psi$ $\in
\sec\mathbf{P}_{\mathrm{Spin}_{1,3}^{e}}(M)\times_{\rho}\mathbb{C}^{4}$
the bilinear covariants are the following sections of
 the exterior algebra bundle of \textit{multivector} fields
 \cite{moro}:
\begin{align}
\sigma &  =\psi^{\dagger}\gamma_{0}\psi,\quad\mathbf{J}=J_{\mu}\mathbf{e}%
^{\mu}=\psi^{\dagger}\gamma_{0}\gamma_{\mu}\psi\mathbf{e}^{\mu},\quad
\mathbf{S}=S_{\mu\nu}\mathbf{e}^{\mu\nu}=\frac{1}{2}\psi^{\dagger}\gamma
_{0}i\gamma_{\mu\nu}\psi\mathbf{e}^{\mu}\wedge\mathbf{e}^{\nu},\nonumber\\
\mathbf{K} &  =\psi^{\dagger}\gamma_{0}i\gamma_{0123}\gamma_{\mu}%
\psi\mathbf{e}^{\mu},\quad\omega=-\psi^{\dagger}\gamma_{0}\gamma_{0123}%
\psi.\label{fierz}%
\end{align}
The set $\{\gamma_{\mu}\}$ refers to the Dirac matrices in chiral
representation (see Eq.(\ref{dirac matrices})). Also $
\{\mathbf{1}_{4},\gamma_{\mu},\gamma_{\mu}\gamma_{\nu},\gamma_{\mu}\gamma
_{\nu}\gamma_{\rho},\gamma_{0}\gamma_{1}\gamma_{2}\gamma_{3}\}$
($\mu,\nu,\rho=0,1,2,3$, and $\mu<\nu<\rho$) is a basis for
$\mathbb{C}(4)$  satisfying
 \cite{lou1}
$\gamma_{\mu}\gamma_{\nu}+\gamma_{\nu}\gamma_{\mu}
=2\eta_{\mu\nu} \mathbf{1}_{4}$ and the Clifford product  is
denoted by juxtaposition. More details on  notations can be found
in \cite{moro,rod}.

Given a fixed spin frame the bilinear covariants are considered
as being the following \textit{operator} fields, for each $x\in M$, as mappings $\mathbb{C}^{4}%
\rightarrow\mathbb{C}^{4}$:
\begin{align}
\sigma &
=\psi^{\dagger}\gamma_{0}\psi,\quad\mathbf{J}=J_{\mu}\gamma^{\mu
}=\psi^{\dagger}\gamma_{0}\gamma_{\mu}\psi\gamma^{\mu},\quad\mathbf{S}%
=S_{\mu\nu}\gamma^{\mu\nu}=\frac{1}{2}\psi^{\dagger}\gamma_{0}i\gamma_{\mu\nu
}\psi\gamma^{\mu\nu},\nonumber\\
\quad\mathbf{K}  &  =\psi^{\dagger}\gamma_{0}i\gamma_{0123}\gamma_{\mu}%
\psi\gamma^{\mu},\quad\omega=-\psi^{\dagger}\gamma_{0}\gamma_{0123}\psi.
\label{11}%
\end{align}\noindent In the case of the electron, described by Dirac spinor
fields (classes 1, 2 and 3 below), $\mathbf{J}$ is a
future-oriented timelike current vector which gives the current of
probability, the bivector $\mathbf{S}$ is associated with the
distribution of intrinsic angular momentum, and the spacelike
vector $\mathbf{K}$ is associated with the direction of the
electron spin. For a detailed discussion concerning such entities,
their relationships and physical interpretation, and
generalizations, see, e.g., \cite{cra,lou1,lou2,holl,hol}.

The bilinear covariants satisfy the Fierz identities
\cite{cra,lou1,lou2,holl,hol}
\begin{equation}
\mathbf{J}^{2}=\omega^{2}+\sigma^{2},\quad\mathbf{K}^{2}=-\mathbf{J}^{2}%
,\quad\mathbf{J}\llcorner\mathbf{K}=0,\quad\mathbf{J}\wedge\mathbf{K}%
=-(\omega+\sigma\gamma_{0123})\mathbf{S}. \label{fi}%
\end{equation}
\noindent A spinor field such that \emph{not both} $\omega$ and
$\sigma$ are null is said to be regular. When $\omega=0=\sigma$, a
spinor field is said to be \textit{singular}.

Lounesto spinor field classification is given by the following
spinor field classes \cite{lou1,lou2}, where in the first three
classes it is implicit that $\mathbf{J}$\textbf{,
}$\mathbf{K}$\textbf{, }$\mathbf{S}$ $\neq0$:

\begin{itemize}
\item[1)] $\sigma\neq0,\;\;\; \omega\neq0$.

\item[2)] $\sigma\neq0,\;\;\; \omega= 0$.\label{dirac1}

\item[3)] $\sigma= 0, \;\;\;\omega\neq0$.\label{dirac2}

\item[4)] $\sigma= 0 = \omega, \;\;\;\mathbf{K}\neq0,\;\;\;
\mathbf{S}\neq0$.

\item[5)] $\sigma= 0 = \omega, \;\;\;\mathbf{K}= 0,
\;\;\;\mathbf{S}\neq0$.\label{elko1}

\item[6)] $\sigma= 0 = \omega, \;\;\; \mathbf{K}\neq0, \;\;\;
\mathbf{S} = 0$.
\end{itemize}

\noindent The current density $\mathbf{J}$ is always non-zero.
Types-(1), -(2), and -(3) spinor fields are denominated
\textit{Dirac spinor fields} for spin-1/2 particles and types-(4),
-(5), and -(6) are respectively called \textit{flag-dipole}
\cite{hopf}, \textit{flagpole}\footnote{ Such spinor fields are
constructed by a null 1-form field current and an also null 2-form
field angular momentum, the ``flag'' \cite{koso}.} and
\textit{Weyl spinor fields}. Majorana spinor fields are a
particular case of a type-(5) spinor field. It is worthwhile to
point out a peculiar feature of types-(4), -(5) and -(6) spinor
fields: although $\mathbf{J}$ is always non-zero,
$\mathbf{J}^{2}=-\mathbf{K}^{2}=0$. It shall be seen below that
the bilinear covariants related to an ELKO spinor field, satisfy
$\sigma=0=\omega,\;\;\mathbf{K}=0,\;\;\mathbf{S}\neq0$ and
$\mathbf{J}^{2}=0$. Since Lounesto proved that there are
\textit{no} other classes based on distinctions among bilinear
covariants, ELKO spinor fields must belong to one of the disjoint
six classes.

Types-(1), -(2) and -(3) Dirac spinor fields (DSFs) have different
algebraic and geometrical characters, and we would like to
emphasize the main differing points. For more details, see e.g.
\cite{lou1,lou2}. Recall that if the quantities $P = \sigma + {\bf
J} + \g_{0123}\omega$ and $Q = {\bf S} + {\bf K}\g_{0123}$ are
defined \cite{lou1,lou2}, in type-(1) DSF we have $P = -(\omega +
\sigma\g_{0123})^{-1}{\bf K}Q$ and also $\psi = -i(\omega +
\sigma\g_{0123})^{-1}\psi$. In type-(2) DSF, $P$ is a multiple of
$\frac{1}{2\sigma} (\sigma + {\bf J})$ and looks like a proper
energy projection operator, commuting with the spin projector
operator given by $\frac{1}{2} (1 - i\g_{0123}{\bf K}/\sigma)$.
Also, $P = \g_{0123}{\bf K}Q/\sigma$. Further, in type-(3) DSF,
$P^2 = 0$ and $P = {\bf K}Q/\omega$. The introduction of the
spin-Clifford bundle makes it possible to consider all the
geometric and algebraic objects
---
 the Clifford bundle, spinor fields, differential form fields, operators and
Clifford fields --- as being elements of an unique unified
formalism. It is well known that spinor fields have three
different, although equivalent, definitions: the operatorial, the
classical and the algebraic one. In particular, the operatorial
definition allows us to factor --- up to sign --- the DSF $\psi$
as $\psi = (\sigma + \omega\g_{0123})^{-1/2}R$, where $R \in$
Spin$^{e}_{1,3}$. Denoting ${\bf K}_k = \psi\g_k\tilde\psi$, where
$\tilde\psi$ denotes the reversion of $\psi$, the set $\{{\bf J},{
K}_1,{ K}_2,{K}_3\}$ is an orthogonal basis of $\RR^{1,3}$. On the
other hand, in classes (4), (5) and (6) --- where $\sigma =
\bar\psi\psi = 0 = \omega = \bar\psi\g_5\psi$, the vectors $\{{\bf
J},{ K}_1,{ K}_2,{ K}_3\}$ no longer form a basis and collapse
into a null-line \cite{lou1,lou2}. In such case only the boundary
term is non null. Finally, to a Weyl spinor field $\xi$ (type-(6))
with bilinear covariants {\bf J} and {\bf K},
 two Majorana spinor fields $\psi_\pm =
\frac{1}{2}(\xi + C(\xi))$ can be associated, where $C$ denotes
the charge conjugation operator. Penrose flagpoles are implicitly
defined by the equation $\sigma + {\bf J} + i{\bf S}-
i\g_{0123}{\bf K}+ \g_{0123}\omega = \frac{1}{2} ({\bf J} \mp
i{\bf S}\g_{0123})$ \cite{lou1,lou2}. For a physically useful
discussion regarding the disjoint classes -(5) and -(6) see, e.g.,
\cite{plaga}. The fact that two Majorana spinor fields $\psi_\pm$
can be written in terms of a Weyl type-(6) spinor field $\psi_\pm
= \frac{1}{2}(\xi + C(\xi))$, is an `accident' when the
(Lorentzian) spacetime has $n= 4$ --- the present case --- or
$n=6$ dimensions. The more general assertion concerns the property
that two Majorana, and more generally ELKO
 spinor fields $\psi_\pm$ can be written in terms of a \emph{pure spinor} field --- hereon denoted
by  $\mmu$ --- as $\psi_\pm = \frac{1}{2}(\mmu + C(\mmu))$. It is
well known that Weyl spinor fields are pure spinor fields when
$n=4$ and $n=6$. When the complexification of $\CC\otimes
\RR^{1,3}$ of $\RR^{1,3}$ is considered, one can consider a
maximal totally isotropic subspace $N$
  of $\CC^{1,3}$, by the Witt decomposition, where $\dim_\CC N = 2$.
Pure spinors are defined by the property  $x\mmu = 0$ for all $x\in
N \subset \CC^{1,3}$. In this context, Penrose flags can be
defined by the expression Re$(i\mmu\tilde\mmu)$.


\begin{thebibliography}{99}

\bibitem{allu}D. V. Ahluwalia-Khalilova and D. Grumiller, \emph{Spin Half
Fermions, with Mass Dimension One: Theory, Phenomenology, and Dark
Matter}, \textit{JCAP} \textbf{07} (2005) 012
[\texttt{arXiv:hep-th/0412080v3}].

\bibitem{alu2} D. V. Ahluwalia-Khalilova and D. Grumiller,
\emph{Dark matter: A spin one half fermion field with mass
dimension one?}, Phys. Rev. {\bf D72} (2005) 067701  [{\tt
arXiv:hep-th/0410192v2}].


\bibitem{gau} D. V. Ahluwalia, Cheng-Yang Lee, D. Schritt, T. F. Watson, {}{%
Dark matter and dark gauge fields}, in ``Dark matter in astroparticle and
particle physics, DARK 2007, Proceedings of the 6th international Heidelberg
conference'' (24-28 September 2007, Sydney, Australia), Eds. H. V.
Klapdor-Kleingrothaus and G. F. Lewis, pp. 198-208. [\texttt{%
arXiv:0712.4190v2 [hep-ph]}]; {}{Local fermionic dark matter with mass
dimension one}, [\texttt{arXiv:0804.1854v4 [hep-th]}].


\bibitem{osmano} R. da Rocha and J. M. Hoff da Silva, \emph{From Dirac spinor fields to eigenspinoren des ladungskonjugationsoperators}, J. Math. Phys. {\bf 48} (2007) 123517 [{\tt arXiv:0711.1103
[math-ph]}].

\bibitem{wal} R. da Rocha and W. A. Rodrigues, Jr., \emph{Where are ELKO spinor fields in Lounesto spinor field classification?}, Mod. Phys. Lett. {\bf A21} (2006) 65-74  [{\tt arXiv:math-ph/0506075v3}].





\bibitem {lou1}P. Lounesto, \emph{Clifford Algebras, Relativity and Quantum
Mechanics}, in P. Letelier and W. A. Rodrigues, Jr. (eds.),
\emph{Gravitation: the Spacetime Structure}, Proc. of the
$8^{\mathrm{th}}$ Latin American Symposium on Relativity and
Gravitation, \'{A}guas de Lind\'{o}ia, Brazil, 25-30 July 1993,
World-Scientific, London 1993.

\bibitem{lou2}P. Lounesto, \emph{Clifford Algebras and Spinors},
2$^{\mathrm{nd}}$ ed., pp. 152-173, Cambridge Univ. Press,
Cambridge 2002.


\bibitem{alu3} D. V. Ahluwalia-Khalilova,
\emph{Dark matter, and its darkness},
    Int. J. Mod. Phys. {\bf D15} (2006) 2267-2278  [{\tt arxiv:hep-th/0603545v3}].


\bibitem{boe1} C. G. Boehmer, {}{The Einstein-Elko system -- Can dark matter
drive inflation?}, \emph{Annalen Phys.} \textbf{16} (2007) 325-341 [\texttt{%
arXiv:gr-qc/0701087v1}]; {}{The Einstein-Cartan-Elko system}, \emph{Annalen
Phys.} \textbf{16} (2007) 38-44 [\texttt{arXiv:gr-qc/0607088v1}]; {}{Dark
spinor inflation -- theory primer and dynamics}, \emph{Phys. Rev.} \textbf{D
77} (2008) 123535 [\texttt{arXiv:0804.0616v1 [astro-ph]}]. 

\bibitem{hopf} R. da Rocha and J. M. Hoff da Silva, \emph{ELKO, flagpole and flag-dipole spinor fields, and the instanton Hopf fibration},  accepted for publication in Adv. Appl. Clifford Alg. (2009) [{\tt arXiv:0811.2717v1 [math-ph]}].

\bibitem{moro} R. A. Mosna and W. A. Rodrigues, Jr., \emph{The bundles of
algebraic and Dirac-Hestenes spinor fields}, J. Math. Phys.
\textbf{45} (2004) 2945-2988 [\texttt{arXiv:math-ph/0212033v5}].


\bibitem{nossoijg} R. da Rocha and J. M. Hoff da Silva, \emph{ELKO Spinor Fields: Lagrangians for Gravity derived from Supergravity}, 
accepted for publication in
 Int. J. Geom. Meth. Mod. Phys. (2009) [{\tt 	arXiv:0901.0883v1 [math-ph]}].
 
 
\bibitem{tung2} R. S. Tung, \emph{Gravitation as a supersymmetric gauge theory},  \emph{Phys. Lett.} {\bf A 264} (2000) 341-345 [{\tt arXiv:gr-qc/9904008}].

\bibitem{ijmpdze} R. da Rocha R and J. G. Pereira,
\emph{The quadratic spinor Lagrangian, axial torsion current, and
generalizations}, \emph{Int. J. Mod. Phys.} {\bf D 16} (2007) 1653-1667
[{\tt arXiv:gr-qc/0703076v1}].
 


\bibitem{rod} W. A. Rodrigues, Jr., \emph{Algebraic and Dirac Hestenes
Spinors and Spinor Fields}, {J. Math. Phys}. \textbf{45} (2004) 2908-2966 [%
\texttt{arXiv:math-ph/0212030v6.}]


\bibitem{cra} J. P. Crawford, \emph{On the Algebra of Dirac Bispinor
Densities: Factorization and Inversion Theorems}, J. Math. Phys.
\textbf{26} (1985) 1429-1441; \textit{The geometric structure of
the space of fermionic physical observables}, em Micali A et
al.(eds.) \textit{Clifford Algebras and Their Applications in
Math. Physics}, Kluwer Acad. Publishers, Dordrecht 1989.

\bibitem{holl} P. R. Holland, \emph{Relativistic Algebraic Spinors and
Quantum Motions in Phase Space}, {Found. Phys.} \textbf{16} (1986)
708-709.

\bibitem{hol} P. R. Holland, \emph{\ Minimal Ideals and Clifford Algebras in
the Phase Space Representation of spin-1/2 Fields}, p. 273-283 in
Chisholm J S R and Common A K (eds.), \emph{Proceedings of the
Workshop on Clifford Algebras and their Applications in
Mathematical Physics (Canterbury 1985)}, Reidel, Dordrecht 1986.


\bibitem{koso} M. R. Francis and A. Kosowsky, \emph{The construction of
spinors in geometric algebra}, Annals Phys. \textbf{317} (2005) 383-409 [%
\texttt{arXiv:math-ph/0403040v2}].


\bibitem{plaga} R. Plaga, \emph{The non-equivalence of Weyl and Majorana
neutrinos with standard-model gauge interactions}, [\texttt{%
arXiv:hep-ph/0108052v1}].


\end{thebibliography}
\end{document}